\documentclass{elsart}
\usepackage{epsfig}
\def\be{\begin{equation}}
\def\ee{\end{equation}}
\def\bea{\begin{eqnarray}}
\def\eea{\end{eqnarray}}

\def\del{\partial}

\def\fr{\frac}
\begin{document}
\begin{flushright}
Cavendish-HEP-00/09\\
DAMTP-2000-90\\
hep-ph/0008229
\\
\vspace*{1cm}
\end{flushright}
\begin{frontmatter}
\hyphenation{}
\boldmath
\title{Gribov's Confinement Scenario\,$^{1,2}$}
\unboldmath
\author{Carlo Ewerz}
\footnotetext[1]{Talk presented at the X.\ Int.\ Light-Cone Meeting on 
Non-Perturbative QCD and Hadron Phenomenology, `From Hadrons to Strings', 
Heidelberg, June 2000}
\footnotetext[2]{Work supported in part by the EU Fourth Framework Programme
`Training and Mobility of Researchers', Network `Quantum Chromodynamics
and the Deep Structure of Elementary Particles',
contract FMRX-CT98-0194 (DG 12 - MIHT).}
\stepcounter{footnote}
\stepcounter{footnote}
\address{Cavendish Laboratory, Cambridge University, \\
Madingley Road, Cambridge CB3 0HE, UK\\
and\\
DAMTP, Centre for Mathematical Sciences, Cambridge University,\\ 
Wilberforce Road, Cambridge CB3 0WA, UK\\
carlo@hep.phy.cam.ac.uk}
\date{August 2000}
\begin{abstract}
I give a brief account of Gribov's 
scenario of supercritical charges in QCD. 
Gribov's equation for the Green function of light quarks 
and its derivation from the corresponding 
Dyson-Schwinger equation are described. 
The resulting Green function is shown to exhibit 
chiral symmetry breaking. 
\end{abstract}
\end{frontmatter}

\section{Introduction}
The confinement of quarks and the breaking of chiral symmetry 
are among the most striking consequences of Quantum Chromodynamics. 
In this talk I give a brief account of two interesting 
suggestions towards their understanding that have been 
made by V.\,N.\ Gribov \cite{Gribov1,Gribov2}. 
The first is the idea that confinement 
is based on the supercritical binding of light quarks. The 
second is a new approach to the difficult problem of solving 
the exact Dyson--Schwinger equation for the quark's 
Green function. This new method has been used to study in detail 
the breaking of chiral symmetry and the analytic structure 
of the light quark's Green function \cite{myself}. 

\section{Confinement as a supercritical phenomenon}
\label{mechanism}
Supercritical charges are well known in QED. An isolated 
pointlike nucleus with a charge $Z>137$ is unstable and 
captures an electron from the vacuum to form a supercritical 
bound state while a positron is emitted. 
This process continues until the charge reaches a subcritical 
value. 
The phenomenon of supercritical charges is possible only 
due to the existence of light fermions making pair creation 
in the strong field energetically possible. 
Gribov's idea is that a similar phenomenon causes 
the confinement of quarks in QCD. In this scenario each 
color charge is supercritical due to the existence of 
very light (almost massless) quarks in our world. 
Contrary to QED this applies even to the color charge 
of a single quark. The vacuum structure of 
light quarks is drastically modified due to this process, 
and the quark becomes a resonance that cannot be 
observed as an asymptotic state. (For a more detailed 
discussion the reader is referred to \cite{Gribov1,myself}.) 
It is expected that the interesting physical picture of 
supercritical charges manifests itself in the Green function 
of the light quark to which we now turn. 

\section{Gribov's equation for the Green function of light quarks}
\label{equation}
In \cite{Gribov1} a new equation for the Green function 
$G(q)$ of a light quark has been suggested. 
In order to derive it one starts from the corresponding 
Dyson--Schwinger integral equation (see for example \cite{DSE}). 
It involves the Green function of the gluon which in 
Feynman gauge is 
$D_{\mu\nu}(k) = - g_{\mu\nu}\alpha_s(k)/k^2$. 
We assume that the effective strong coupling 
constant $\alpha_s(k)$ is a slowly varying function 
and stays finite at low momenta. This assumption is in 
agreement with results obtained recently in the dispersive 
approach to power corrections, see for example \cite{BPY}. 
The Feynman gauge is unique in this context because it allows 
one to obtain an especially simple equation for $G(q)$. 
Its characteristic factor $1/q^2$ yields a four--dimensional 
delta--function $\delta^{(4)}(q)$ under the action of 
the d'Alembert operator $\del^2=\del^\mu \del_\mu$, 
where $\del_\mu=\del/\del q^\mu$. 
After applying $\del^2$ to the Dyson--Schwinger equation 
one can identify the most singular contributions -- namely 
the delta--functions -- to the integrals from the infrared region 
where the confining dynamics resides. 
Collecting only these most singular terms\footnote{This 
method can in principle 
be extended to include also subleading terms.} 
one can perform the integral, and the resulting terms can 
be shown to involve two factors representing full vertex functions. 
Using Ward identities the latter can in turn be expressed in 
terms of derivatives of $G(q)$ itself. We are then left with 
a differential equation for the Green function of light 
quarks in Feynman gauge, 
\be
\del^2 G^{-1} = C_F \fr{\alpha_s(q)}{\pi} 
\,(\del^\mu G^{-1}) \,G \,(\del_\mu G^{-1}) 
\,,
\label{gribgl}
\ee
where $C_F=4/3$. 
The running coupling $\alpha_s(q)$ ensures that 
the equation reproduces at large spacelike momenta 
the correct mass and wave function renormalization 
known from perturbation theory. 

\section{Chiral symmetry breaking and critical coupling}
\begin{figure}
\begin{center}
\input{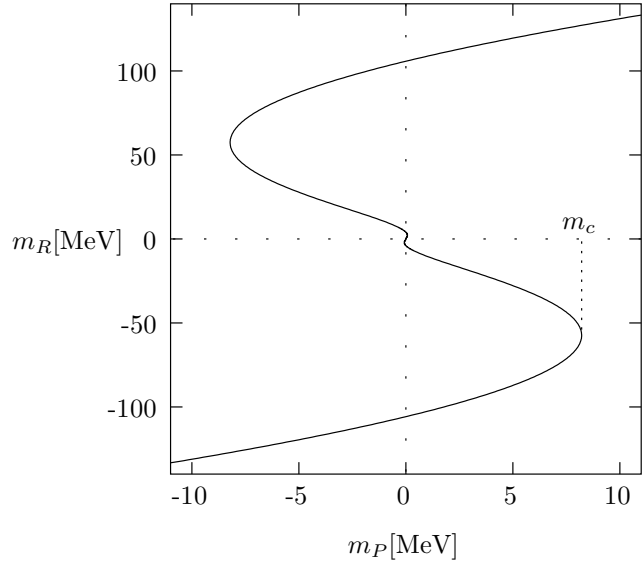}
\caption{\label{fig:m2m}
Relation between perturbative mass $m_P$ and renormalized 
mass $m_R$ 
(obtained with a model for $\alpha_s$ motivated by 
\protect\cite{BPY})
}
\end{center}
\end{figure}

Eq.\ (\ref{gribgl}) can be used to study the behaviour 
of the Green function in different limits analytically. 
A full numerical study was performed in \cite{myself}. 
We will now restrict ourselves to discussing the dynamical mass 
function $M(q^2)$ of the light quark in the region of space--like 
momenta, $q^2<0$. 
It turns out that the Green function shows a characteristically 
different behaviour if the effective strong coupling 
constant $\alpha_s (q)$ exceeds a critical value in some 
interval of the momentum $q$ in the infrared region. 
The critical coupling is found to be 
\be
\label{critcoupl}
\alpha_c = \frac{\pi}{C_F} \left( 1- \sqrt{\fr{2}{3}}\,\right)
\simeq 0.43
\,.
\ee
This different behaviour can be shown to correspond to the 
breaking of chiral symmetry. In order to see this 
we define a `perturbative' mass $m_P=M(\lambda^2)$ 
at some large scale $\lambda$ where perturbation 
theory holds. The perturbative mass is, roughly speaking, 
similar to the current mass of the quark. Let us further 
define the `renormalized' mass $m_R=M(0)$ as the 
low--momentum limit of the dynamical mass function. 
If the coupling $\alpha_s$ remains subcritical the relation 
between $m_P$ and $m_R$ is monotonic, and $m_R$ 
vanishes for vanishing $m_P$. If the coupling becomes larger 
than $\alpha_c$ this picture is drastically changed, 
and the resulting relation between $m_P$ and $m_R$ is 
shown in fig.\ \ref{fig:m2m}. 
We see that the renormalized mass remains finite even for 
vanishing perturbative mass, and chiral symmetry is thus 
broken. 

The breaking of chiral symmetry has been discussed in the 
context of supercritical charges also in \cite{Kiev}, and 
similar values for the critical coupling have been obtained 
from different approximations to the Dyson--Schwinger equations, 
see \cite{DSE}. 
A unique feature of the approach outlined above is that 
eq.\ (\ref{gribgl}) is a differential equation which makes 
it technically very simple to study its solutions in the 
whole complex momentum plane. 

\section{Outlook}
It turns out that the analytic structure of the Green function 
obtained from eq.\ (\ref{gribgl}) exhibits poles and cuts 
at positive $q^2$ and does therefore not correspond to confined 
quarks \cite{myself}. This was expected \cite{Gribov1} 
since this equation does not take into account the effects 
of pions which arise as Goldstone bosons. 
The special r\^ole of pions in the scenario of supercritical 
charges suggests that they be included as elementary 
degrees of freedom and that eq.\ (\ref{gribgl}) be modified 
accordingly \cite{Gribov2}. 
There are indications that the modified equation does 
in fact lead to a confining Green function, but a full analysis 
still remains to be done. 


%

\begin{thebibliography}{99}
\bibitem{Gribov1}
V.\,N.\ Gribov, {\it Eur.\ Phys.\ J. }{\bf C 10} (1999) 71 
[hep-ph/9807224]
\bibitem{Gribov2}
V.\,N.\ Gribov, {\it Eur.\ Phys.\ J. }{\bf C 10} (1999) 91
[hep-ph/9902279]
\bibitem{myself}
C.\ Ewerz,  {\it Eur.\ Phys.\ J. }{\bf C 13} (2000) 503 
[hep-ph/0001038]
\bibitem{DSE}
C.\,D.\ Roberts, A.\,G.\ Williams, 
{\it Prog.\ Part.\ Nucl.\ Phys.\ }{\bf 33} (1994) 477 
[hep-ph/9403224] 
\bibitem{BPY}
Yu.\,L.\ Dokshitzer, G.\ Marchesini, B.\,R.\ Webber, 
{\it Nucl.\ Phys.\ }{\bf B 469} (1996) 93 [hep-ph/9512336] 
\bibitem{Kiev}
P.\,I.\ Fomin, V.\,P.\ Gusynin, V.\,A.\ Miransky, Yu.\,A.\ Sitenko, 
Riv.\ Nuovo Cim.\ {\bf 6} No 5 (1983) 1 and references therein
\end{thebibliography}
\end{document}